\documentclass[useAMS,usenatbib,usegraphicx,twocolumn]{mn2e}

\usepackage{epsfig}
\usepackage{times}
\def\apj{ApJ}
\def\apjs{ApJS}
\def\araa{ARAA}
\def\mnras{MNRAS}

\def\aj{AJ}
\def\apjl{ApJL}

\def\aap{Astronomy \& Astrophysics}

\title[Disk Surface Density and Halo Spin Parameter]{How Does the Surface Density and Size of Disk Galaxies Measured in Hydrodynamic Simulations Correlate with the Halo Spin Parameter?}
\author[J. Kim and J. Lee]{Ji-hoon Kim$^{1}$\thanks{E-mail: me@jihoonkim.org} and Jounghun Lee$^{2}$\thanks{E-mail: jounghun@astro.snu.ac.kr}\\
$^{1}$Department of Astronomy and Astrophysics, University of California, Santa Cruz, CA 95064, USA\\
$^{2}$Astronomy Program, Department of Physics and Astronomy, Seoul National University, Seoul 151-747, Korea}
\begin{document}

\date{Submitted 2013 January 23}

\pagerange{\pageref{firstpage}--\pageref{lastpage}} \pubyear{2013}

\maketitle

\label{firstpage}

\begin{abstract}
Late-type low surface brightness galaxies (LSBs) are faint disk galaxies with central maximum stellar surface densities below $100\,\,M_{\odot}\,{\rm pc}^{-2}$. 
The currently favored scenario for their origin is that LSBs have formed in fast-rotating halos with large angular momenta. 
We present the first numerical evidence for this scenario using a suite of self-consistent hydrodynamic simulations of a $2.3\times 10^{11}\,M_{\odot}$ galactic halo, in which we investigate the correlations between the disk stellar/gas surface densities and the spin parameter of its host halo.  
A clear anti-correlation between the surface densities and the halo spin parameter $\lambda$ is found. 
That is, as the halo spin parameter increases, the disk cutoff radius at which the stellar surface density drops below $0.1\,\,M_{\odot}\,{\rm pc}^{-2}$ monotonically increases, while the average stellar surface density of the disk within that radius decreases. 
The ratio of the average stellar surface density for the case of $\lambda=0.03$ to that for the case of $\lambda=0.14$ reaches more than 15.  
We demonstrate that the result is robust against variations in the baryon fraction, confirming that the angular momentum of the host halo is an important driver for the formation of LSBs.
\end{abstract}


\section{INTRODUCTION}\label{sec:intro}

Despite years of effort, the origin and true nature of dark matter are still shrouded in mystery.  
Astrophysical interest in dark matter has been escalated not only because it is inferred to dominate the matter mass budget of the Universe, but because of the simplicity of its dynamics.  
In particular, dark matter is thought to not only exist inside galactic-scale halos, but also heavily influence the galactic gravitational dynamics.  
Naturally, the records of dark matter dynamics throughout the formation of the galactic halos are imprinted in the morphology of the present-day galaxies.   
We may therefore advance our understanding of the gravity and the nature of dark matter by investigating the galaxies' structure and formation history.  
For this purpose, late-type low surface brightness galaxies (LSBs) provide us with a unique and intriguing laboratory. 

Late-type LSBs typically refer to the disk galaxies that are fainter than the night sky, exhibiting $B$-band central surface brightnesses $\mu_{B, \,0} > 22.7\,\, {\rm mag}\,\,{\rm arcsec}^{-2}$ \citep{freeman70} or equivalently central maximum stellar surface densities $\Sigma_{\star, \,0} <100\,\,M_{\odot}\,{\rm pc}^{-2}$ \citep{mcgaugh-etal01}. 
They are different from the under-luminous dwarf spheroidals in that the late-type LSBs may have large disks and high intrinsic luminosities in spite of their low surface brightnesses.
Their abundance has long been underestimated, but observations now imply that LSBs may represent a significant fraction of the galaxy population \citep[e.g.][and references therein]{IB97, OB00}. 
Since it was realized that a wealth of LSBs inhabit both cluster and field regions, and they may provide an important clue for the galaxy formation process \citep[e.g.][]{impey-etal88,irwin-etal90,mcgaugh-etal95,sprayberry-etal96, sprayberry-etal97}, many studies have explored their physical properties, dynamical states, host halo structures, spatial distribution, and environmental dependences \citep[e.g.][]{MB94, mo-etal94, zwaan-etal95, deblok-etal96, mihos-etal96, gerritsen-etal99, bell-etal00, MW01, bergmann-etal03, kuzio-etal04, matthews-etal05, boissier-etal08, rosenbaum-etal09, gao-etal10, galaz-etal11, morelli-etal12, zhong-etal12, ceccarelli-etal12}.

\begin{figure*}
\begin{center}
\includegraphics[width=5.4in]{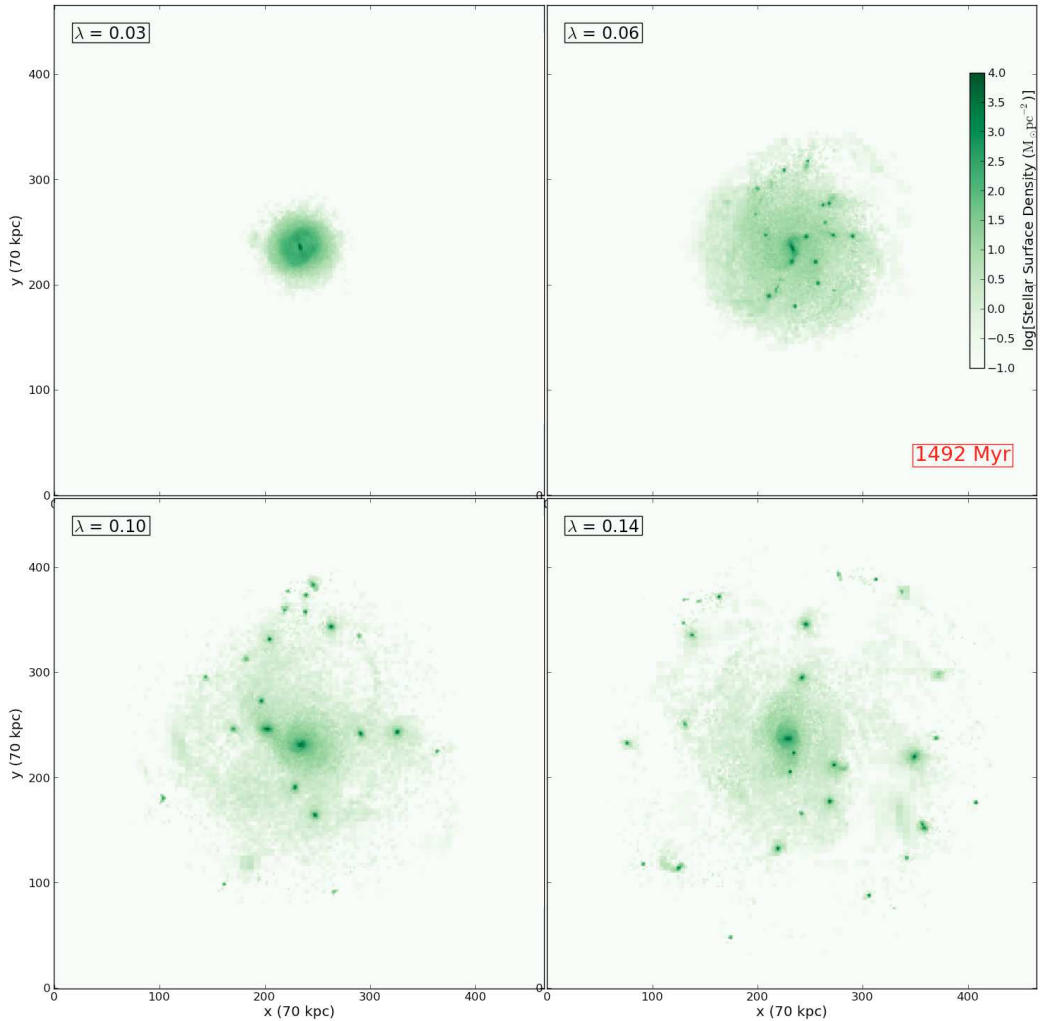}
\caption{Face-on stellar surface densities in a central 70 kpc box at 1.49 Gyr after the start of the simulation for four different halo spin parameters, $\lambda=0.03, \, 0.06, \, 0.10,\, 0.14$ in the {\it top-left, top-right, bottom-left, and bottom-right} panel, respectively.  The initial baryonic mass fraction in the halo is $f_{\rm b}=0.10$.  Stars of age less than 1.0 Gyr are used to estimate surface densities.  More information on the suite of simulations and analysis is provided in \S\ref{sec:sim} and \S\ref{sec:density}.}
\label{fig:faceon_snapshot}
\end{center}
\end{figure*}

\begin{figure*}
\begin{center}
\includegraphics[width=5.4in]{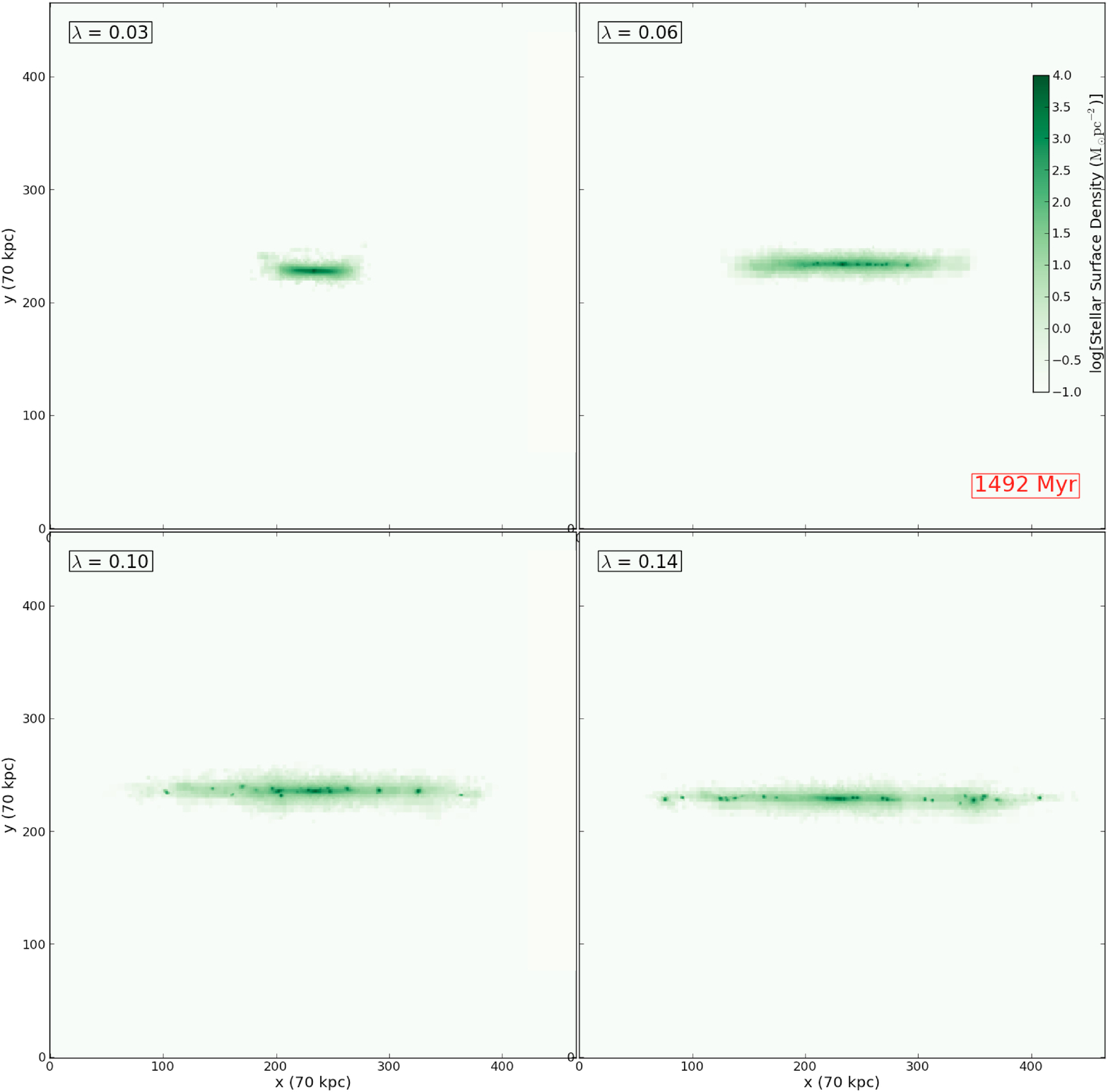}
\caption{Same as Figure \ref{fig:faceon_snapshot} but from the edge-on angle.  $\lambda=0.03, \, 0.06, \, 0.10,\, 0.14$ in the {\it top-left, top-right, bottom-left, and bottom-right} panel, respectively.}
\label{fig:edgeon_snapshot}
\end{center}
\end{figure*}

While the origin of LSBs is yet to be fully understood, theoretical modelings to comprehend how LSBs have formed paralleled the above phenomenological efforts. 
The simplest, but currently favored scenario states that the difference of LSBs from the high surface brightness galaxies (HSBs) results from the fast-spinning motion of their host halos \citep{dalcanton-etal97,jimenez-etal97,mo-etal98}. 
To quantitatively compare the predictions of this theory with the observed properties of LSBs, \citet{jimenez-etal98} utilized simplified models for the physical, chemical, and spectrophotometric evolution of disk galaxies, and found that the theory nicely reproduces the observed properties of LSBs. 
Their results were later confirmed by \citet{boissier-etal03} who used larger and higher quality LSB samples (see \S\ref{sec:review} for more discussion). 

Because LSBs are more dark matter dominant than HSBs \citep[e.g.][]{BM97,pickering-etal97,mcgaugh-etal01}, LSBs have also been employed to test the nature of dark matter and gravity.
Studies have claimed that the rotation curves of LSBs could be useful in constraining the properties of dark matter and even modified Newtonian dynamics \citep[e.g.][]{MB98a,MB98b,BM98,SS00,swaters-etal10,KS11}.
Recently \citet{lee-etal12} speculated that the abundance of LSBs could be a testbed for modified gravity.
Using a high-resolution simulation, they demonstrated that modified gravity may spin up the low mass galactic halos and significantly alter the abundance of LSBs. 
{\it Nevertheless}, in order to utilize LSBs as a probe into the nature of dark matter and gravity, it is indispensable to first properly constrain the driver that produces LSBs.  
Although the galactic evolution models of \citet{jimenez-etal98} and \citet{boissier-etal03} supported the hypothesis that the LSBs form in the halos with large angular momenta, their results were subject to the simplified prescriptions for the formation of disk galaxies. 
To improve the state of the field and advance our understanding of the origin of LSBs, a comprehensive, {\it self-consistent} hydrodynamic simulation is greatly needed.\footnotemark
\footnotetext{Readers should note that a few other scenarios of the origin of LSBs have been tested in numerical simulations, including the one through disk instability \citep[e.g.][]{noguchi01} and the one through collisional formation of a ring galaxy \citep[e.g.][]{mapelli-etal08}.}

In light of this development, we for the first time numerically examine whether or not the angular momentum of the host halo is an important driver for determining the galactic surface brightness. 
Using a suite of hydrodynamic simulations, we quantitatively investigate the correlations between the surface density of a disk galaxy and the spin parameter of its host halo without having to rely on a simplified prescription for galactic evolution. 
The outline of this article is as follows. 
We briefly describe the currently popular model for the origin of LSBs in \S\ref{sec:review}, and the set-up of our numerical experiments in \S\ref{sec:sim}.  
The simulation results is presented and analyzed in \S\ref{sec:results} focusing on the correlation between the stellar/gas surface densities and the halo spin parameter.  
The robustness of our result is also examined against variations in the initial baryonic mass fraction.
Finally we discuss the implications and limitations of our results and draw a conclusion in \S\ref{sec:discuss}.

\section{A BRIEF REVIEW OF THE MODEL FOR THE ORIGIN OF LOW SURFACE BRIGHTNESS GALAXIES}\label{sec:review}

We first briefly review the currently popular model for the origin of low surface brightness galaxies, which is put to the numerical test in the present article.  

A dark matter halo acquires its angular momentum ${\bf J}$ through the tidal interactions with the surrounding matter distribution at its proto-halo stage \citep{peebles69, dor70, white84}. 
Since the magnitude of the halo angular momentum is dependent upon its virial mass $M_{\rm vir}$, it is often convenient to adopt a dimensionless spin parameter $\lambda \equiv |{\bf J}|/(2GM_{\rm vir}R_{\rm vir})^{1/2}$ with the halo virial radius $R_{\rm vir}$ to quantify how fast a halo rotates around its axis of symmetry. 
A series of cosmological $N$-body simulations have been employed to investigate the structural properties of dark matter haloes, including the spin parameter \citep[e.g.][]{bullock-etal01, avila-reese-etal05, maccio-etal07}.
These simulations have shown that the probability density distribution of $\lambda$ is approximated by a log-normal distribution in $\Lambda$CDM cosmology, being relatively insensitive to mass scale, environment, and redshift \citep[e.g.][]{bullock-etal01}.

In the gravitationally self-consistent model for the formation of disk galaxies, a rotationally-supported disk forms in a potential well of a dark matter halo through a dissipative gravitational collapse of the baryonic content  \citep{FE80}. 
The resulting baryonic disk shares the same tidally-induced specific angular momentum {\bf j} (angular momentum per unit mass) with its host halo.  
Based on this model, \citet{dalcanton-etal97} proposed a scenario that the high angular momentum disk in a dark matter halo of large {\bf j} have its baryonic mass spread over a wider extent, leading to a large disk cutoff radius and thus a low {\it gas} surface density \citep[see also][]{jimenez-etal97,mo-etal98}.
To relate the gas surface density to the {\it stellar} surface density and to perform a systematic test of this scenario against observations, \citet{jimenez-etal98} computed various galactic properties including surface brightnesses by adopting the following prescriptions for galaxy evolution:  {\it (a)} an isothermal spherical halo, {\it (b)} a model for gas infall rate that reproduces the observed properties of the Milky Way, {\it (c)} the Schmidt law for star formation \citep{schmidt59}, and {\it (d)} the galactic chemical evolution model by \citet{MF89}.  
They found nice agreement between the resulting predictions and the observed properties of LSBs such as surface brightness, metallicity, and colors, provided that the host halo satisfies the condition of $\lambda \ge 0.06$. 
This work was extended by \citet{boissier-etal03} to a larger set of observational data and a wider range of halo spin parameters.  
They drew the same conclusion that a LSB is likely to form in a dark matter halo with a large spin parameter. 
They also noted that some modulation in the star formation rate history would be desired to achieve a better agreement between the theory and observations.
In similar fashion, \cite{avila-reese-etal05} employed a semianalytic model of disk galaxy evolution to demonstrate the dependence of disk properties on its environment.

However, the conclusion reached by these studies is contingent upon how the gas infall rate and star formation rate were prescribed in their overly simplified evolution models. 
It is still inconclusive whether the theoretical model by \citet{dalcanton-etal97} would hold robustly in reality where the gas infall and star formation can hardly be specified by a simple analytic recipe. 
In what follows, we describe a suite of hydrodynamic simulations with radiative gas cooling and star formation of a galactic halo with varying spin parameters, designed to better address this long-standing problem. 
Most importantly, unlike previous studies, we do not rely on simplified prescriptions for star formation and gas infall rates.
This enables us to examine the robustness of the theory in a simplified, yet the most realistic set-up until now.

\begin{figure}
\begin{center}
\includegraphics[width=3.5in]{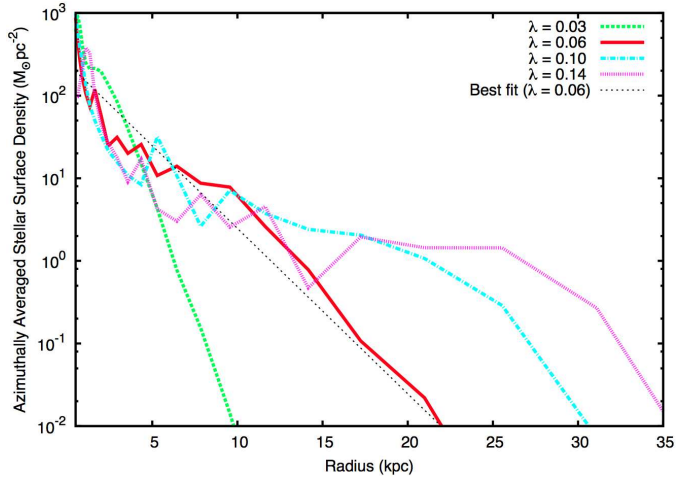}
\caption{Azimuthally-averaged stellar surface densities at 1.49 Gyr after the start of the simulation with $f_{\rm b}=0.10$ for $\lambda=0.03,\, 0.06,\, 0.10,\, 0.14$ as a {\it dashed, solid, dot-dashed, and dotted line}, respectively.  The cutoff radius $R_{\rm c}$ at which the stellar surface density falls below $0.1\,\,M_{\odot}\,{\rm pc}^{-2}$ monotonically increases as the halo spin parameter increases (see Table \ref{tab:scale_radius} and Figure \ref{fig:spin_vs_cutoff_radius}).  The {\it thin dotted line} shows the best fit of the disk stellar surface density for $\lambda=0.06$ to an exponential profile.}
\label{fig:star_profile}
\end{center}
\end{figure}

\section{INITIAL CONDITIONS AND PHYSICS IN THE CODE}\label{sec:sim}

We simulate the evolution of a dwarf-sized galactic halo with varying halo spin parameters and initial baryon fractions.  
The initial conditions of our experiment and the physics included in the code are explained in this section. 

We set up an isolated halo of total mass $2.3\times10^{11}\,M_{\odot}$ with four halo spin parameters, $\lambda = 0.03,\, 0.06,\, 0.10,\, 0.14$, and with two different initial baryon fractions for each $\lambda$,  $ f_{\rm b} = (M_{\star} + M_{\rm gas}) / M_{\rm total} = 0.05$ and 0.10 (see Table \ref{tab:scale_radius}; 95\% dark matter + 5\% gas for $f_{\rm b} = 0.05$, and 90\% dark matter + 10\% gas for $f_{\rm b} = 0.10$). 
We first create a dataset of $10^6$ collisionless particles, to which we add gas by splitting the particles (with an initial metalliticy of $0.003\,\,Z_{\odot}$).  
Gas and dark matter follow the same shapes of a Navarro-Frenk-White profile \citep{nfw97} with concentration $c = 10$, and have the same value of $\lambda$ as an averaged circular motion.  
This way, as the gas cools down, the progenitor forms a disk galaxy embedded in a gaseous halo.  
Note that this procedure is very different from that of \citet{jimenez-etal98} where a galactic disk already exists when their calculation starts. 
For detailed descriptions to set up the initial condition, see \cite{kim-etal12}. 

We then follow the evolution of each galactic halo in a 1 ${\rm Mpc}^3$ box for $\sim 3$ Gyr using the ZEUS hydrodynamics solver included in the publicly available adaptive mesh refinement {\it Enzo-2.1}.\footnote{http://enzo-project.org/} 
Multi-species chemistry (H, ${\rm H}^+$, He, ${\rm He}^+$, ${\rm He}^{++}$, ${\rm e}^-$) and non-equilibrium cooling modules are employed to calculate radiative losses. 
Cooling by metals is estimated in gas above $10^4$ K with the rates tabulated by \cite{SD93}, and below $10^4$ K by \cite{GJ07}.
We refine the cells by factors of two in each axis on the overdensities of gas and dark matter.  
In the finest cell of size $\Delta\, x = 122 \,\,{\rm pc}$, a star cluster particle is produced with initial mass $M_{\rm sc}^{\rm init} = 0.5 \rho_{\rm gas} \Delta \,x^3$ when {\it (a)} the proton number density exceeds 3.2 ${\rm cm^{-3}}$, {\it (b)} the velocity flow is converging, {\it (c)} the cooling time $t_{\rm cool}$ is shorter than the dynamical time $t_{\rm dyn}$ of the cell, and {\it (d)} the particle produced has at least $104000 \,\,M_{\odot}$.
For each star cluster particle, $1.5 \times 10^{-6}$ of its rest mass energy and 20\% of its mass are returned to the gas phase over 12 $t_{\rm dyn}$.  
This represents various types of stellar feedback, chiefly the injection of thermal energy by supernova explosion.  
More information on adopted baryonic physics can be found in \cite{kim-etal09} and \cite{kim-etal11}. 

\begin{figure}
\begin{center}
\includegraphics[width=3.5in]{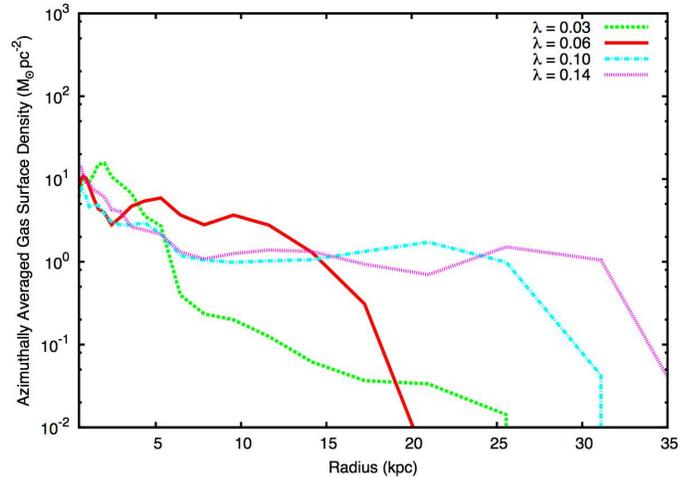}
\caption{Same as Figure \ref{fig:gas_profile} but for azimuthally-averaged gas surface densities.  The size of a gas disk increases with the halo spin parameter.}
\label{fig:gas_profile}
\end{center}
\end{figure}

\begin{table*}  
\begin{minipage}{150mm}
\caption{Halo spin parameter, disk cutoff/scale radius, and average stellar surface density. For definitions of variables, see \S\ref{sec:density}.}
\begin{tabular}{ c  ||  c c c  |  c c c } 
\hline\hline   
 & \multicolumn{3}{c |}{$f_{\rm b}=0.05$} & \multicolumn{3}{c}{$f_{\rm b}=0.10$}  \\ [0.5 ex]
$\lambda$  & 
cutoff $R_{\rm c}$ [kpc] & scale $R_{\rm s}$ [kpc] & $\Sigma_{\star, {\rm ave}}$ [$M_{\odot} \,{\rm pc}^{-2} $]  & 
cutoff $R_{\rm c}$ [kpc] & scale $R_{\rm s}$ [kpc] & $\Sigma_{\star, {\rm ave}}$ [$M_{\odot} \,{\rm pc}^{-2} $]  \\ [1ex] 
\hline      
0.03    & 7.2   & 0.92 & 157.0  & 8.1    & 0.82 & 278.2\\   
0.06    & 13.3 & 1.7   & 49.1    & 17.3  & 2.2   & 60.2\\
0.10    & 26.0 & 3.3   & 11.7    & 26.8  & 3.6   & 24.6\\
0.14    & 32.2 & 4.4   & 6.5      & 32.5  & 4.9   & 15.2\\  [0.5ex] 
\hline
\end{tabular} 
\label{tab:scale_radius}
\end{minipage}
\end{table*}

\begin{figure}
\begin{center}
\includegraphics[width=3.65in]{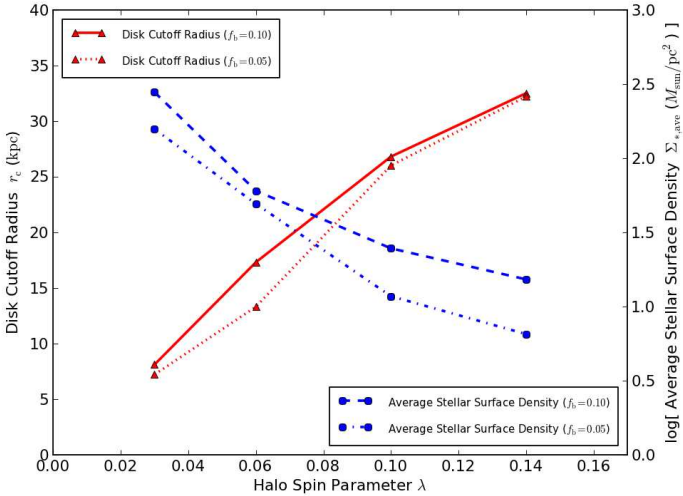}
\caption{Correlation between the halo spin parameter $\lambda$ and the disk cutoff radius $R_{\rm c}$ ({\it triangles, solid} and {\it dotted lines}) and average stellar surface density $\Sigma_{\star, {\rm ave}}$ ({\it circles, long dashed} and {\it dot-dashed lines}), for two different simulations of baryon fraction $f_{\rm b}=0.05$ and 0.10 at 1.49 Gyr after they start.}
\label{fig:spin_vs_cutoff_radius}
\end{center}
\end{figure}

\section{SIMULATION RESULTS}\label{sec:results}

We now describe the results of our simulations with varying halo spin parameters and baryon fractions.  
First, we focus on the correlations between surface densities of a galactic disk and the spin parameter of its host halo in the simulations with an initial baryonic mass fraction of 10\%. 
 
\subsection{Correlation Between Halo Spin Parameter and Size of Galactic Disk}\label{sec:density}

We start by calculating the stellar surface density of the simulated disk galaxy for each case of $\lambda$ at 1.49 Gyr after the start of the simulation with an initial baryon fraction $f_{\rm b} = 0.10$, the ratio of baryonic mass to total mass.
We choose to use the snapshot at 1.49 Gyr, the midpoint of our simulation for $\sim3$ Gyr.\footnote{Our choice to analyze the epoch of 1.49 Gyr is to avoid both the first $<$ 0.5 Gyr of artificial starbursts driven by the initially unstable gas profile, and the last $<$ 1.0 Gyr when the galactic halo is devoid of star-forming gas due to the unrealistic lack of cosmological inflows.  While the qualitative conclusion of our analysis is not significantly affected by the choice of the epoch, we reason that the 1.49 Gyr snapshot best represents the reported simulation, rendering realistic values of stellar-to-gas ratio in the halo ($\sim$ 75\% of the gas turned into stars in the simulation with $f_{\rm b} = 0.10$ and $\lambda=0.06$).}
Shown in Figures \ref{fig:faceon_snapshot} and \ref{fig:edgeon_snapshot} are the snapshots of face-on and edge-on stellar surface densities using the stars of age less than 1.0 Gyr.  
In each figure, the four panels correspond to the four different values of halo spin parameters, $\lambda = 0.03,\, 0.06,\, 0.10,\, 0.14$. 
The snapshots are generated in a 70 kpc box with uniform $150$ pc resolution which can be regarded as an equivalent of an aperture size in photometric observations. 
We also point out that, in these figures and the subsequent analysis, we exclude the stars formed in the first 0.49 Gyr since they would most possibly reflect the artificial burst of star formation in the initially unimpeded collapse, caused by an unstable gas profile at the beginning of the simulation.\footnote{Excluding the stars formed in the first 0.49 Gyr we, therefore, implicitly assume that the stars younger than 1 Gyr form the disk of a dwarf-sized galaxy.  While 1 Gyr may correspond to an interval between major disturbances for dwarf-sized galaxies, our choice is certainly fiducial.}
From these figures, it is clearly noticeable that the size of the stellar disk is larger in the case of a higher value of $\lambda$, indicating that the large angular momentum of a disk in a dark matter halo of large {\bf j} have the baryonic mass spread to a wider extent.

\begin{figure}
\begin{center}
\includegraphics[width=3.33in]{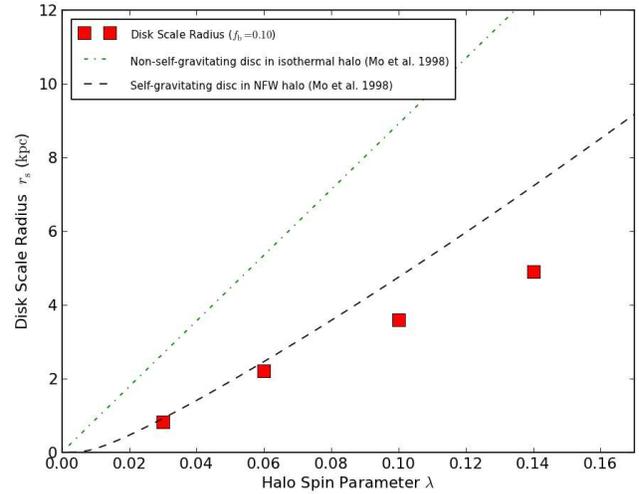}
\caption{Correlation between the halo spin parameter $\lambda$ and the disk scale radius $R_{\rm s}$ at 1.49 Gyr after the start of the simulation with $f_{\rm b}=0.10$.  Also shown are the analytic predictions from \citet{mo-etal98} for a non-self-gravitating disk in an isothermal halo ({\it dot-dahsed line}) and a self-gravitating disk in a NFW halo \citep[{\it long dashed line};][]{nfw97}.}
\label{fig:spin_vs_scale_radius}
\end{center}
\end{figure}

To quantify the change in the disk size resulting from different halo spin parameters, the radial profile of the {\it stellar} surface density, $\Sigma_{\star}(r)$ (in the unit of $M_{\odot}\,{\rm  pc}^{-2}$), is determined by taking the azimuthal average of the stellar density on the disk plane at 1.49 Gyr after the start of the simulation.
The result of this exercise is displayed in Figure \ref{fig:star_profile}. 
Similarly, the radial profile of the azimuthally-averaged {\it gas} surface density, $\Sigma_{\rm gas}(r)$, is shown in Figure \ref{fig:gas_profile}.
The disk cutoff radius, $R_{\rm c}$, is then defined as the radius at which the stellar surface density drops below $0.1\,\,M_{\odot}\,{\rm pc}^{-2}$. 
Figure \ref{fig:star_profile} also shows the best fit of the stellar surface density of the galaxy with $\lambda=0.06$ to an exponential profile, $\Sigma_{\star}(r) \sim e^{-r/R_{\rm s}}$ (thin dotted line).
In this way the scale radius of a disk, $R_{\rm s}$, is defined as the radius at which the density on this best fit curve is reduced to 1/$e$ times the central maximum.
Table \ref{tab:scale_radius} compiles $R_{\rm c}$ and $R_{\rm s}$ for four different $\lambda$'s, and the average stellar surface densities within the cutoff radius, $\Sigma_{\star, {\rm ave}}$.
Meanwhile Figure \ref{fig:spin_vs_cutoff_radius} visualizes the relationship between $\lambda$, $R_{\rm c}$, and $\Sigma_{\star, {\rm ave}}$.

\begin{figure}
\begin{center}
\includegraphics[width=3.45in]{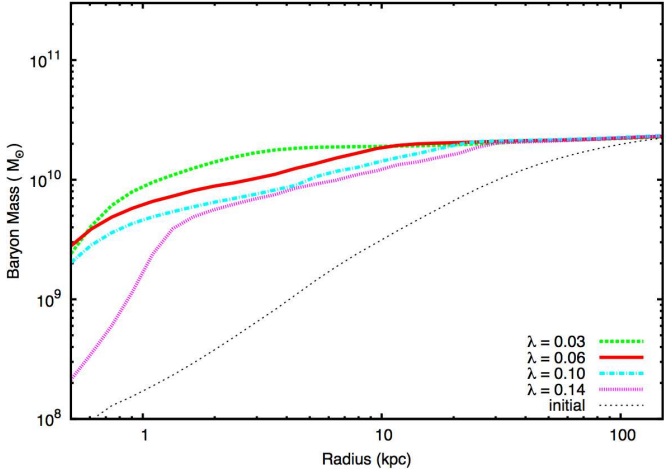}
\caption{Radial profiles of enclosed baryonic mass (stars+gas) at 1.49 Gyr after the start of the simulation with $f_{\rm b}=0.10$ for $\lambda=0.03,\, 0.06,\, 0.10,\, 0.14$ as a thick dashed, solid, dot-dashed, and dotted line, respectively.  The initial radial profile is also shown for comparison as a thin dotted line.  While all the halos show the results of dissipative collapses towards the center, the halo of $\lambda=0.03$ has the densest core in the baryon distribution.}
\label{fig:bmass_profile}
\end{center}
\end{figure}

In Figure \ref{fig:spin_vs_cutoff_radius} one can immediately observe that the cutoff radius $R_{\rm c}$  monotonically increases as $\lambda$ increases, from 8.1 kpc at $\lambda = 0.03$ to 32.5 kpc at $\lambda = 0.14$. 
The net effects are an enlarged photometric size of the stellar disk, and a lowered average stellar surface density  $\Sigma_{\star, {\rm ave}}$:  $278.2\,\,M_{\odot}\,{\rm pc}^{-2}$ at $\lambda = 0.03$ to $15.2\,\,M_{\odot}\,{\rm pc}^{-2}$ at $\lambda = 0.14$. 
As a result, the ratio of $\Sigma_{\star, {\rm ave}}$ for the case of $\lambda=0.03$ to that of $\lambda=0.14$ becomes larger than 15.  
This simplified, yet self-consistent numerical calculation validates the scenario that the disk in a faster-rotating halo have its star-forming gas extend over a larger area thanks to the stabilization by the higher angular momentum barrier.  
It then entails a lower stellar surface density averaged on the disk. 
When a uniform value of stellar mass to light ratio is assumed across the set of simulated galaxies, the low stellar surface density is translated into the low surface brightness, giving results consistent with the previous studies \citep{jimenez-etal98, boissier-etal03, avila-reese-etal05}.

It is also informative to compare our result with the analytic predictions by \citet{mo-etal98}.
In Figure \ref{fig:spin_vs_scale_radius} the disk scale radii, $R_{\rm s}$, of the run with $f_{\rm b}=0.10$ are drawn, along with the analytic formulae for the cases of {\it (a)} a non-self-gravitating disk embedded in an isothermal halo,\footnote{$R_{\rm s}(\lambda) = 89.1\,\lambda$ kpc from Eq.(12) of \citet{mo-etal98}, with $r_{200} = 126$ kpc for a $2.3\times10^{11}M_{\odot}$ halo and $j_{\rm d} = m_{\rm d} = 0.1$} and {\it (b)} a self-gravitating disk embedded in a more realistic Navarro-Frenk-White halo \citep[NFW;][]{nfw97}.\footnote{$R_{\rm s}(\lambda) = 47.5\,\lambda(\lambda/0.1)^{0.211+0.0047/\lambda}$ kpc from Eq.(28) of \citet{mo-etal98}, with $r_{200} = 126$ kpc, $c = 10$, and $j_{\rm d} = m_{\rm d} = 0.1$}
Obviously the scale radius $R_{\rm s}$ in our simulation monotonically increases with $\lambda$, from 0.92 kpc at $\lambda = 0.03$ to 4.4 kpc at $\lambda = 0.14$.
One could also notice that the agreement between our numerical result and the \citet{mo-etal98} prediction using a self-gravitating disk in a NFW halo is surprisingly good despite many simplifications and idealized assumptions adopted in their analytic derivation (e.g. no baryonic physics or dissipation is considered).

\begin{figure}
\begin{center}
\includegraphics[width=3.45in]{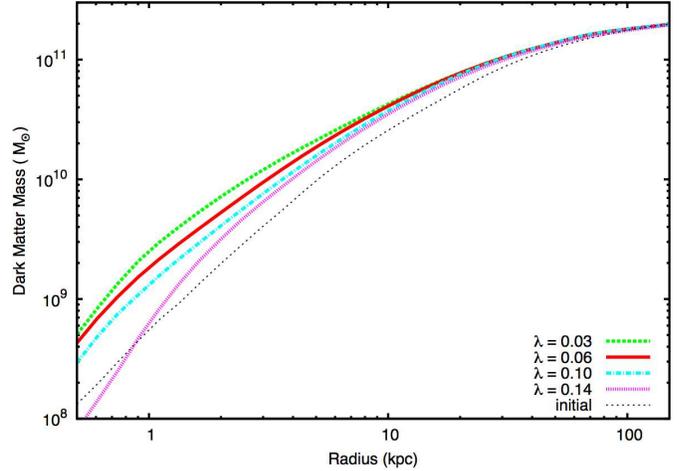}
\caption{Same as Figure \ref{fig:bmass_profile} but of the enclosed dark matter mass.  Note that the halo of $\lambda=0.03$ exhibits the most centrally-concentrated density profile.}
\label{fig:DMmass_profile}
\end{center}
\end{figure}

\begin{figure}
\begin{center}
\includegraphics[width=3.5in]{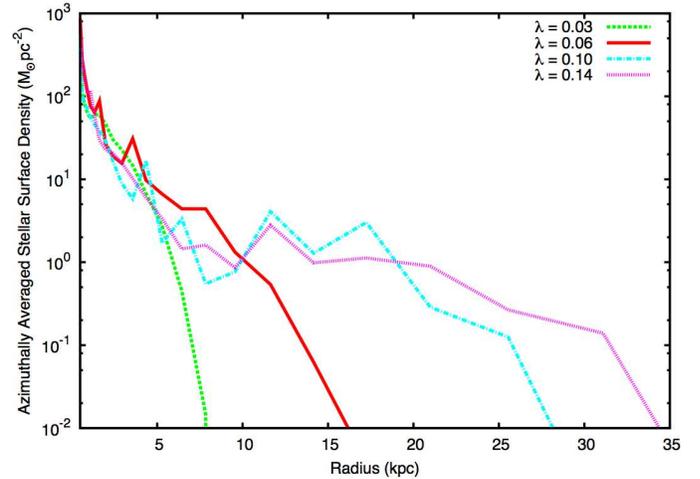}
\caption{Same as Figure \ref{fig:star_profile} but for the case of the initial baryonic mass fraction $f_{\rm b}=0.05$ instead of $f_{\rm b}=0.10$.  The trend of an increasing cutoff radius with a larger halo spin parameter persists.}
\label{fig:star_profile2}
\end{center}
\end{figure}

In addition, we investigate how the enclosed mass profiles are correlated with the halo spin parameter.  
We evaluate the profiles of baryonic mass enclosed within a sphere of radius $r$ from the galactic center of mass at 1.49 Gyr after the start of the simulation.
The resulting radial profiles in Figure \ref{fig:bmass_profile} allow us to cleanly see the discrepancy across different runs in a cumulative manner. 
While all the halos show the results of dissipative collapses towards the center, the halo of the lowest spin parameter $\lambda=0.03$ has the densest core in the baryonic distribution, enclosing most of the baryons within $\sim$ 3 kpc from the center. 
We also inspect the half-mass radius that encloses a half of the total baryonic mass inside the halo, $\sim2.2\times 10^{10}\, M_{\odot}$. 
As is anticipated, the half-mass radius increases almost monotonically with $\lambda$, from 1.4 kpc at $\lambda = 0.03$ to 8.0 kpc at $\lambda = 0.14$. 
Finally Figure \ref{fig:DMmass_profile} demonstrates the profiles of the dark matter mass enclosed within a sphere of radius $r$. 
One can easily notice that the larger the halo spin parameter is, the less dense the dark matter distribution becomes in the core region ($r \la$ 1 kpc). 
In particular, the mass enclosed within a 1 kpc sphere in the $\lambda = 0.03$ case is $\sim 2.8\times 10^9\, M_{\odot}$, whereas that of $\lambda = 0.14$ case is only $\sim 6.0\times 10^8\, M_{\odot}$.  
This implies that dark matter itself also gradually collapses inward unless additionally supported by the angular momentum.

\subsection{Robustness Against Variations in the Baryon Fraction}\label{sec:fbar}

Now that we have found a strong correlation between the halo spin parameter and the stellar/gas surface densities of a galactic disk, it will be very illuminating to examine whether or not this correlation is robust against variations in the initial baryonic mass fraction $f_{\rm b}$. 
This examination may also allow us to determine which factor is more dominant in determining the disk size and galactic surface density, between the absolute amount of gas and the angular momentum of the host halo.  
Recall that the results presented in \S\ref{sec:density} are all obtained by the simulations with $f_{\rm b}=0.10$.    

We here repeat the same analysis in Figure \ref{fig:star_profile} but on a suite of simulations with a different initial baryon fraction, $f_{\rm b}=0.05$, and plot the result in Figure \ref{fig:star_profile2}. 
First, as is obviously expected, a decreased value of $f_{\rm b}$ also reduces the overall stellar surface density for a halo with the same spin parameter. 
What is more intriguing is that the positive correlation between the disk cutoff radius $R_{\rm c}$ and the halo spin parameter $\lambda$ still holds strong in simulations with a lower $f_{\rm b}$: a faster rotation of the host halo necessitates a larger cutoff radius of the stellar disk formed inside.  
Table \ref{tab:scale_radius} and Figure \ref{fig:spin_vs_cutoff_radius} again shows the cutoff radii, scale radii, and the average stellar surface densities of the galactic disks within those radii for $f_{\rm b}=0.05$. 
It again clearly displays that for both $f_{\rm b}$ the correlations between $R_{\rm c}$ (or $R_{\rm s}$) and $\lambda$ hold the same.  
Interestingly, the difference between two $R_{\rm c}$'s for $f_{\rm b}=0.05$ and 0.10 with the same $\lambda$ is surprisingly small.  
For example, the cutoff radius of a disk in the halo of $\lambda=0.10$ and $f_{\rm b}=0.10$ is 26.8 kpc, only slightly expanded from 26.0 kpc in the halo of the same $\lambda$ but of $f_{\rm b}=0.05$.  
This is despite the fact that the initial gas supply is boosted by a factor of two.  
Although it should be considered provisional this finding may suggest that, between $\lambda$ and $f_{\rm b}$, the more influential factor for determining the cutoff radius is $\lambda$, the spin parameter of the halo.

\subsection{Notes on the Performed Runs}\label{sec:note}

Before we proceed to conclude, it is worth to note a few points on the simulations we have described so far.  
We draw the reader's attention to the central maximum stellar surface densities of the simulated galaxies in Figure \ref{fig:star_profile}, which still remain above or very close to $100\,\,M_{\odot}\,{\rm pc}^{-2}$ regardless of $\lambda$.  
Therefore, at face value, even the largest halo spin parameter seemingly fails to produce a LSB galaxy in its conventional definition (see \S\ref{sec:intro}).
{\it However}, these simulations are under a particular assumption of a relatively high baryon fraction of $f_{\rm b}=0.10$. 
By comparing Figures \ref{fig:star_profile} and \ref{fig:star_profile2}, one observes that the central peak stellar surface density indeed drops significantly when the initial gas supply is reduced by a factor of two. 
Had the galaxy been simulated in a fast-rotating halo (e.g. $\lambda > 0.10$) with an even smaller baryon fraction (e.g. $f_{\rm b} < 0.02$), it would likely have developed into a LSB galaxy classically defined with a central maximum stellar surface density less than $100\,\,M_{\odot}\,{\rm pc}^{-2}$.
Further, by running simulations in an isolated set-up, we also have made an implicit assumption that the halo does not experience any interaction with its surrounding environment, such as mergers and harassment.  
Including such effects might have drastically changed the surface densities of the simulated galaxies.  

In summary, the reported simulations are aimed to make a comparison between the galaxies that differ only by their halo spin parameters.  
Our experiment, however, is {\it not} specifically designed to turn {\it any} galaxy into a LSB with a large halo spin parameter.  
In reality the galactic surface brightness is most likely determined by the combination of a number of  factors such as the halo angular momentum, halo density profile, baryon fraction, gas infall rate, stellar population and feedback, and environmental effects.
These are just a few constituents in a large multi-dimensional parameter space left to be explored.

\section{DISCUSSION AND CONCLUSION}\label{sec:discuss}

We have presented the first numerical evidence for the hypothesis that late-type LSBs form in the halos with large angular momenta.  
A suite of hydrodynamic simulations with radiative gas cooling and star formation of a $2.3\times 10^{11}\,M_{\odot}$ galactic halo is employed with various halo spin parameters and baryon fractions.
We have investigated the correlations between the stellar/gas surface densities of a galactic disk and the spin parameter of its host halo.  
A clear signal of anti-correlation is found between the surface densities and the halo spin parameter. 
That is, as the halo spin parameter increases, the cutoff radius, defined as the radius at which the stellar surface density drops below $0.1\,\,M_{\odot}\,{\rm pc}^{-2}$, monotonically increases, while the average stellar surface density of the disk within that radius decreases. 
The ratio of the average stellar surface density for the case of $\lambda=0.03$ to that for the case of $\lambda=0.14$ is larger than 15.   
We also have demonstrated that the result is robust against variations in the initial value of the baryonic mass fraction.  

Our simplified, yet self-consistent numerical experiment confirms that one of the most important drivers for the formation of LSBs is the spin parameters of their host halos.  
This investigation provides a critical link between observations and the theoretical models for the origin of LSBs, which has been missing in the previous studies.  
It also enables us to better comprehend the formation process of LSBs, and to utilize LSBs for the purpose of constraining the nature of invisible component that dominates galactic dynamics.  
Even so, the reported calculation alone {\it cannot} decide whether the halo angular momentum is the most determinitive factor in the formation of LSBs.  
Further, the limitations of the test should be noted as this experiment in an idealized, undisturbed halo does not resolve the subgrid physics below the simulation resolution. 
A more realistic description of subgrid physics such as star formation and feedback \citep{guedes-etal11, agertz-etal12, kim-etal12, hopkins-etal13}, joined with higher numerical resolution, will need to be considered in the future in the context of hierarchical structure formation.

\section*{Acknowledgments}

J. K. thanks Mark Krumholz and the anonymous referee for providing insightful comments and stimulating advice on the earlier version of this article.  
J. K. gratefully acknowledges support from the NSF Grant AST-0955300.  
J. L. acknowledges the financial support from the National Research Foundation of Korea (NRF) grant funded by the Korea government (MEST, No. 2012-0004916) and from the National Research Foundation of Korea to the Center for Galaxy Evolution Research (No. 2010-0027910). 
The examination and analysis of the simulation data are greatly aided by an AMR analysis toolkit {\it yt} \citep{turk-etal11}.

\label{lastpage}


\begin{thebibliography}{}
\bibitem[Agertz et al.(2012)]{agertz-etal12} 
Agertz, O., et al.\ 2012, arXiv:1210.4957
\bibitem[Avila-Reese et al.(2005)]{avila-reese-etal05} 
Avila-Reese, V., et al.\ 2005, \apj, 634, 51
\bibitem[Bell et al.(2000)]{bell-etal00} Bell, E.~F., Barnaby, D., 
Bower, R.~G., et al.\ 2000, \mnras, 312, 470 
\bibitem[Bergmann et al.(2003)]{bergmann-etal03} Bergmann, M.~P., 
J{\o}rgensen, I., \& Hill, G.~J.\ 2003, \aj, 125, 116 
\bibitem[Boissier et al.(2003)]{boissier-etal03} Boissier, S., 
Monnier-Ragaigne, D., Prantzos, N., et al.\ 2003, \mnras, 343, 653 
\bibitem[Boissier et al.(2008)]{boissier-etal08} Boissier, S., Gil de 
Paz, A., Boselli, A., et al.\ 2008, \apj, 681, 244 
\bibitem[Bullock et al.(2001)]{bullock-etal01} Bullock, J.~S., Dekel, 
A., Kolatt, T.~S., et al.\ 2001, \apj, 555, 240 
\bibitem[Ceccarelli et al.(2012)]{ceccarelli-etal12} Ceccarelli, L., 
Herrera-Camus, R., Lambas, D.~G., Galaz, G., 
\& Padilla, N.~D.\ 2012, \mnras, L500 
\bibitem[Dalcanton et al.(1997)]{dalcanton-etal97} Dalcanton, J.~J., 
Spergel, D.~N., \& Summers, F.~J.\ 1997, \apj, 482, 659 
\bibitem[de Blok et al.(1996)]{deblok-etal96} de Blok, W.~J.~G., 
McGaugh, S.~S., \& van der Hulst, J.~M.\ 1996, \mnras, 283, 18 
\bibitem[de Blok \& McGaugh(1997)]{BM97} 
de Blok, W.~J.~G., \& McGaugh, S.~S.\ 1997, \mnras, 290, 533 
\bibitem[de Blok \& McGaugh(1998)]{BM98} 
de Blok, W.~J.~G., \& McGaugh, S.~S.\ 1998, \apj, 508, 132 
\bibitem[Doroshkevich(1970)]{dor70} Doroshkevich,
A.~G.\ 1970, Astrofizika, 6, 581
\bibitem[Fall \& Efstathiou(1980)]{FE80} 
Fall, S.~M., \& Efstathiou, G.\ 1980, \mnras, 193, 189 
\bibitem[Freeman(1970)]{freeman70} 
Freeman, K.~C.\ 1970, \apj, 160, 811
\bibitem[Galaz et al.(2011)]{galaz-etal11} Galaz, G., Herrera-Camus, 
R., Garcia-Lambas, D., \& Padilla, N.\ 2011, \apj, 728, 74 
\bibitem[Gao et al.(2010)]{gao-etal10} Gao, D., Liang, Y.-C., Liu, 
S.-F., et al.\ 2010, Research in Astronomy and Astrophysics, 10, 1223 
\bibitem[Gerritsen \& de Blok(1999)]{gerritsen-etal99} 
Gerritsen, J.~P.~E., \& de Blok, W.~J.~G.\ 1999, \aap, 342, 655 
\bibitem[Glover \& Jappsen(2007)]{GJ07}
Glover, S.~C.~O., \& Jappsen, A.\ 2007, \apj, 666, 1
\bibitem[Guedes et al.(2011)]{guedes-etal11} 
Guedes, J., et al.\ 2011 \apj, 742, 76
\bibitem[Hopkins et al.(2013)]{hopkins-etal13} 
Hopkins, P.~F, et al.\ 2013, \mnras, 430, 1901 
\bibitem[Impey et al.(1988)]{impey-etal88} 
Impey, C., Bothun, G., \& Malin, D.\ 1988, \apj, 330, 634 
\bibitem[Impey \& Bothun(1997)]{IB97} 
Impey, C., \& Bothun, G.\ 1997,  \araa, 35, 267 
\bibitem[Irwin et al.(1990)]{irwin-etal90} Irwin, M.~J., Davies, 
J.~I., Disney, M.~J., \& Phillipps, S.\ 1990, \mnras, 245, 289 
\bibitem[Jimenez et al.(1997)]{jimenez-etal97} Jimenez, R., Heavens, 
A.~F., Hawkins, M.~R.~S., \& Padoan, P.\ 1997, \mnras, 292, L5 
\bibitem[Jimenez et al.(1998)]{jimenez-etal98} Jimenez, R., Padoan, 
P., Matteucci, F., \& Heavens, A.~F.\ 1998, \mnras, 299, 123 
\bibitem[Kim et al.(2009)]{kim-etal09} Kim, J.-H., Wise, J.~H., 
\& Abel, T.\ 2009, \apjl, 694, L123 
\bibitem[Kim et al.(2011)]{kim-etal11} Kim, J.-H., Wise, J.~H., 
Alvarez, M.~A., \& Abel, T.\ 2011, \apj, 738, 54 
\bibitem[Kim et al.(2012)]{kim-etal12} Kim, J.-H., Krumholz, M.~R., 
Wise, J.~H., Turk, M.~J., Goldbaum, N.~J., \& Abel, T.\ 2012, arXiv:1210.3361
\bibitem[Kuzio de Naray et al.(2004)]{kuzio-etal04} Kuzio de Naray, 
R., McGaugh, S.~S., \& de Blok, W.~J.~G.\ 2004, \mnras, 355, 887 
\bibitem[Kuzio de Naray \& Spekkens(2011)]{KS11} Kuzio de Naray, R.,
  \& Spekkens, K.\ 2011, \apjl, 741, L29
\bibitem[Lee et al.(2013)]{lee-etal12} Lee, J., Zhao, G.-B., Li, 
B., \& Koyama, K.\ 2013, \apj, 763, 28
\bibitem[Maccio et al.(2007)]{maccio-etal07} 
Maccio, A.~V, et al.\ 2007, \mnras, 378, 55 
\bibitem[Mapelli et al.(2008)]{mapelli-etal08} 
Mapelli, M., et al.\ 2008, \mnras, 383, 1223 
\bibitem[Matteucci \& Francois(1989)]{MF89} 
Matteucci, F., \& Francois, P.\ 1989, \mnras, 239, 885 
\bibitem[Matthews et al.(2005)]{matthews-etal05} 
Matthews, L.~D., Gao, Y., Uson, J.~M., \& Combes, F.\ 2005, \aj, 129, 1849 
\bibitem[Matthews \& Wood(2001)]{MW01} 
Matthews, L.~D., \& Wood, K.\ 2001, \apj, 548, 150 
\bibitem[McGaugh \& Bothun(1994)]{MB94} 
McGaugh, S.~S., \& Bothun, G.~D.\ 1994, \aj, 107, 530
\bibitem[McGaugh et al.(1995)]{mcgaugh-etal95} McGaugh, S.~S., 
Schombert, J.~M., \& Bothun, G.~D.\ 1995, \aj, 109, 2019 
\bibitem[McGaugh \& de Blok(1998a)]{MB98a} 
McGaugh, S.~S., \& de Blok, W.~J.~G.\ 1998, \apj, 499, 41
\bibitem[McGaugh \& de Blok(1998b)]{MB98b} 
McGaugh, S.~S., \& de Blok, W.~J.~G.\ 1998, \apj, 499, 66 
\bibitem[McGaugh et al.(2001)]{mcgaugh-etal01} McGaugh, S.~S., Rubin, 
V.~C., \& de Blok, W.~J.~G.\ 2001, \aj, 122, 2381 
\bibitem[Mihos et al.(1996)]{mihos-etal96} Mihos, C., McGaugh, S., 
\& de Blok, E.\ 1996, arXiv:astro-ph/9612115 
\bibitem[Mo et al.(1994)]{mo-etal94} 
Mo, H.~J., McGaugh, S.~S., \& Bothun, G.~D.\ 1994, \mnras, 267, 129 
\bibitem[Mo et al.(1998)]{mo-etal98} Mo, H.~J., Mao, S., 
\& White, S.~D.~M.\ 1998, \mnras, 295, 319 
\bibitem[Morelli et al.(2012)]{morelli-etal12} Morelli, L., Corsini, 
E.~M., Pizzella, A., et al.\ 2012, \mnras, 423, 962 
\bibitem[Navarro et al.(1997)]{nfw97} 
Navarro, J.~F., Frenk, C.~S., \& White, S.~D.~M.\ 1997, \apj, 490, 493 
\bibitem[Noguchi(2001)]{noguchi01} 
Noguchi, M.\ 2001, \mnras, 328, 353
\bibitem[O'Neil \& Bothun (2000)]{OB00} 
O'Neil, K., \& Bothun, G.\ 2000, \apj, 529, 811 
\bibitem[Peebles(1969)]{peebles69} Peebles, P.~J.~E.\ 1969, \apj, 
155, 39
\bibitem[Pickering et al.(1997)]{pickering-etal97} Pickering, T.~E., 
Impey, C.~D., van Gorkom, J.~H., \& Bothun, G.~D.\ 1997, \aj, 114, 1858 
\bibitem[Rosenbaum et al.(2009)]{rosenbaum-etal09} 
Rosenbaum, S.~D., Krusch, E., Bomans, D.~J., \& Dettmar, R.-J.\ 2009,
\aap, 504, 807 
\bibitem[Schmidt(1959)]{schmidt59} 
Schmidt, M.\ 1959, \apj, 129, 243 
\bibitem[Spergel \& Steinhardt(2000)]{SS00} 
Spergel, D.~N., \& Steinhardt, P.~J.\ 2000, Physical Review Letters, 84, 3760 
\bibitem[Sprayberry et al.(1996)]{sprayberry-etal96} Sprayberry, D., 
Impey, C.~D., \& Irwin, M.~J.\ 1996, \apj, 463, 535 
\bibitem[Sprayberry et al.(1997)]{sprayberry-etal97} Sprayberry, D., 
Impey, C.~D., Irwin, M.~J., \& Bothun, G.~D.\ 1997, \apj, 482, 104 
\bibitem[Sutherland \& Dopita(1993)]{SD93}
Sutherland, R.~S., \& Dopita, M.~A.\ 1993, \apjs, 88, 253
\bibitem[Swaters et al.(2010)]{swaters-etal10} Swaters, R.~A., 
Sanders, R.~H., \& McGaugh, S.~S.\ 2010, \apj, 718, 380 
\bibitem[Turk et al.(2011)]{turk-etal11} 
Turk, M.~J., et al.\ 2011, \apjs, 192, 9
\bibitem[White(1984)]{white84} 
White, S.~D.~M.\ 1984, \apj, 286, 38 
\bibitem[Zhong et al.(2012)]{zhong-etal12} Zhong, G.~H., Liang, 
Y.~C., Liu, F.~S., et al.\ 2012, arXiv:1205.2404 
\bibitem[Zwaan et al.(1995)]{zwaan-etal95} Zwaan, M.~A., van der 
Hulst, J.~M., de Blok, W.~J.~G., \& McGaugh, S.~S.\ 1995, \mnras, 273, L35
\end{thebibliography}
\end{document}